# Accounting for Availability Biases in Information Visualization

Evanthia Dimara [1,2]
evanthia.dimara@gmail.com

Pierre Dragicevic [1]
pierre.dragicevic@inria.fr

Anastasia Bezerianos [2,1]
anastasia.bezerianos@lri.fr

[1]INRIA

[2]Univ Paris-Sud & CNRS (LRI)

## ABSTRACT

The availability heuristic is a strategy that people use to make quick decisions but often lead to systematic errors. We propose three ways that visualization could facilitate unbiased decision-making. First, visualizations can alter the way our memory stores the events for later recall, so as to improve users' long-term intuitions. Second, the known biases could lead to new visualization guidelines. Third, we suggest the design of decision-making tools that are inspired by heuristics, e.g. suggesting intuitive approximations, rather than target to present exhaustive comparisons of all possible outcomes, or automated solutions for choosing decisions.

## Categories and Subject Descriptors

H.5.2 [**Information Interfaces and Presentation**]: User Interfaces - Graphical user interfaces

## 1. INTRODUCTION

We all want to make good decisions. However, decision-making judgments often involve approximate estimations of probabilities and frequencies. In order to reduce the complexity of estimation people rely on a limited number of strategies. One of these strategies is called *Availability Heuristic* [14].

The availability heuristic is a rule of thumb in which decision makers "estimate the frequency or probability by the ease with which instances or associations could be brought to mind"[20]. For example, news about a terrible plane crash may temporarily alter our feelings on flight safety. This heuristic simplifies some otherwise very difficult judgments, and it is usually effective since in principle it is easier to recall or imagine common events than uncommon ones.

However, apart from the actual frequency or probability, other factors affect the ease of recalling instances, and thus estimating frequencies for making decisions. Some of these factors affecting recall, illustrated by Tversky and Kahneman [21], often lead to systematic errors:

- *Bias due to retrievability of instances*: People evaluate the probability of an event as higher, when they retrieve its instances effortlessly. Schwarz et al [16] asked one group of participants to recall 12 examples of their past assertive behavior, and another group to recall only 6. After that, they rated their assertiveness. The 12 examples were harder to be recalled than the 6, so the first group rated themselves as less assertive than the second group. Retrievability is often related to a) familiarity [20] or what Whittlesea [23] refers to as the "illusion of pastness"; b) saliency where one instance elicits more attention than another [18]; and c) recency where for example the serial presentation of information may affect memorization [22].

- *Bias due to the effectiveness of the search set*: The generation of a search set depends also on the performed search task. When we ask to compare the instances of the word 'love' with the word 'door' the first seems more frequent. A main reason for this is that besides the comparison of words, there is a hidden task of recalling contexts in which these words appear. It is generally easier to recall abstract contexts than concrete ones [8].

- *Bias of imaginability*: When the frequency of an instance is not stored in memory, we sometimes generate this frequency according to some rule. For example when we want to estimate which is more frequent, the existence of committees of 8 members or of 2, we will mentally construct committees and rate them by the ease of this construction. The mental construction of 2 member committees is easier, and thus may be considered as most frequent [20]. In real life imaginability biases can lead us to overestimate some risks with vivid scenarios and underestimate dangerous risks that are hard to conceive.

- *Bias due to illusory correlation*: When two events co-occur people tend to overestimate the frequency of natural association. For example it is common to patients with paranoia to have peculiar eyes. This association misled undergraduate clinicians to diagnose as paranoid patients with no other symptoms related to paranoia in their medical data, simply because they were guided by a given picture of the patient with peculiar eyes [4][5].

Availability bias affects the decision-making ability in an unconscious way, and can lead people to irrational decisions. We believe we can better support decision-making, through the design of visualizations that take into account these factors that influence decision-making. We illustrate this in a voting scenario that we imagine takes place with and without hypothetical visualizations designed to account for biases.

## 2. THE VOTING DECISION

Imagine that one needs to decide which political candidate to vote for. There are three steps. As a first step, she shapes an opinion on which are the important personal and society issues based on her past exposure to information (media, social environment, personal experiences etc.). Second, she investigates the candidates' former actions, background and current positions, and estimates their ability and willingness to solve the important society issues. Finally, she compares all the alternatives and decides on a candidate.

In an ideal world, voters are aware of their position in the complex political landscape, understand statistical analysis data and micro-macro economics, and have endless memory capacity and time to process all the relative candidates' history. In reality, voters usually simplify this decision using heuristics. However, as we discussed, the common heuristics like availability may lead one to pick a candidate according to, for example: meaningless actions that media over-cover, without important impact in the society [6]; the sequence of their presentation in the public debate [12]; the vividness of the way they talk [15][13][2] or even whether their victory is an event easy to envision [3].

## 3. INFOVIS ON A COMPLEX DECISION

The actual challenge on the three decision-making steps in the voting problem is how to filter, understand, recall and compare information. In principle, this challenge is related to the infovis objectives. But how could visualizations actually assist a voter to reduce availability biases?

*Visualizations to aid recall:* Let's think of a scenario with simple hypothetical visualizations involved in the voting problem. Consider also two alternative policies: a consolidation of the national health system focusing on cancer cure, and a high-cost terrorism counteraction. People tend to make wrong estimations on most probable causes of death in their country [7], misled by media. In contrast, imagine the voter had access to a map with stacked (men/women) bar charts of death causes (society issues awareness) that she can filter. For the simplicity of the argument here we assume a voter who wants to maximize her own self-interest. Thus, the displayed causes can be filtered according to her family's medical records and other individual characteristics (personal issues awareness). The user interacts with the visualization, composing a view that is focused on her interests. From the amount of information that she was able to process, she captures a snapshot of her self-constructed view of the visualization to save among her personal notes. In the last step of the decision-making, she evaluates the policies of each candidate. Based on her memorable interaction experience with the visualization, she intuitively evaluates which election promise has greater impact in her life.

Visualizations could thus educate decision makers to develop unbiased intuitions of their surroundings. However, as we saw in the previous example, this does not only imply visual comparisons of choices and consequences. The availability heuristic succeeds when memory stores the frequent events in an easy-to-recall way. In our example, the visualization facilitates the user's memory by capturing her self-constructed summary. Thus, visualizations could go beyond simply showing all possible alternatives. Visualizations should also help decision makers easily recall the important take-away information.

*Visualizations to remove biases:* So visualizations can help deal with biases. But visualization designers can also reuse the knowledge of robust biases already studied in psychology literature to make better visualizations. Studies on how the candidates' order in the set of ballot papers affects the vote rank, confirm that "recency effect" [22] occurs also in visualizations [24]. Thus visualizations can reinforce the importance of some information due to some known biases (e.g., presenting them last).

Moreover, if the magnitude of a visual variable does not reflect its real impact, we may reinforce not only visual perception bias (bigger is more important), but also retrievability biases (bigger may be easier to remember). For example, consider the two political candidates suggest either the increase of unemployment allowance, or the tax exemption of families with a lot of children. Both are fair measures, but unemployment applies to a larger part of the population. However, families with many children may not be able to survive with the current tax policy. A visualization where the visual variable depends exclusively on the population size is legible, but may lead voters to evaluate the policies only according to the population criterion, even for voters who have 10 children themselves. Visualizations, in addition to what the media can offer, should be able to also display a customized perspective and alternative views of the data. These customized views of the data will be the ones most likely retrieved from their memory during the decision process.

The visualization design should also take into account the biases due to imaginability. When it comes to radical ideas, the mind's inability to construct the outcome of this idea can lead a person to consider it as impossible to happen. In the voting problem, if a candidate proposes "decentralization of state power to local communities", the voters may reject it not only because they disagree, but also because the outcome is an event hard to envision. A conceptualized, but vivid, map representation of the idea of decentralization could alter the voter's willingness to accept a change.

*Visualizations inspired by heuristics:* We mostly discussed so far how to present the information to lead to effective and unbiased decisions. However, the decision-making process itself can be hard even when all the information we need is available. Automated decision-making tools that give explicit answers according to probability computations, may not always feel intuitive and understandable even by experts [9]. They often restrict users by expecting a very particular input, or ignoring other context-relevant information that the users may have [19]. On the other hand common visualization tools often hide uncertainty in the data [17][1] and do not actually shield decision makers against perceptual and cognitive biases [10][14][11] Visualization tools are currently designed and evaluated based on data retrieval and insight tasks, rather than on the ultimate and crucial task of decision-making [1]. Thus we could drive some inspiration of how heuristic strategies like availability, simplify decision tasks, find the effective tradeoff among simplicity and accuracy, and apply this analogy on new visualization tools. That is, decision tools may need to allow some imperfection for the sake of understandability.

## 4. CONCLUSION

We discussed so far how information visualization could eliminate biases due to availability, when we visualize the actual probabilities, rearrange the sequence of information, increase their saliency or filter a subset of them for later recall. While designing such systems, our goal should not be to eliminate the use of heuristics that can cause these biases altogether, but rather to exploit them helping users make better decisions.

To sum up, we suggest that information visualization can reduce availability biases and assist decision-making in three ways. First,

we can take advantage of the good use of availability heuristics and improve users' long-term intuitions. Second, visualizations could provide design techniques that eliminate the known availability biases. Third, we can investigate new decision-making tools that target to inspired by rather than replace the way that availability heuristics work.